\documentclass{aa}
\usepackage{graphicx}
\begin{document}
\title{NGC 2580 and NGC 2588: \\
Two open clusters in the Third Galactic Quadrant \\
       \thanks{Based on observations collected at ESO, CASLEO and CTIO},
       \thanks{Data is only available in electronic form at the CDS via
               anonymous ftp to {\tt cdsarc.u-strasbg.fr (130.79.128.5)} or
           via {\tt http://cdsweb.u-strasbg.fr/cgi-bin/qcat?J/A+A//}
           }
       }
       \author{{}G. Baume\inst{1,2},
         A. Moitinho\inst{3},
         E. E. Giorgi\inst{2},
         G. Carraro\inst{1}
         \and
         R. A. V\'{a}zquez\inst{2}
          }

   \offprints{baume@pd.astro.it}

   \institute{Dipartimento di Astronomia, Universit\`a di Padova,
              Vicolo Osservatorio 2, I-35122 Padova, Italy
         \and
             Facultad de Ciencias Astron\'omicas y Geof\'{\i}sicas de la
         UNLP, IALP-CONICET, Paseo del Bosque s/n, La Plata, Argentina
     \and
CAAUL, Observat\'orio Astron\'omico de Lisboa, Tapada da Ajuda,
  1349-018 Lisboa, Portugal
         }

   \date{Received **; accepted **}

   \abstract{ We present CCD broad band photometric observations in
     the fields of the Third Galactic Quadrant open clusters NGC~2580
     and NGC~2588 ($V(I)_C$ and $UBV(RI)_C$ respectively). From the
     analysis of our data we found that NGC~2580 is located at a
     distance of about 4 kpc and its age is close to 160~Myr. As for
     NGC~2588, it is placed at about 5 kpc from the Sun and is 450~Myr
     old. This means that NGC~2588 belongs to the extension of the 
     Perseus arm, whereas NGC~2580 is closer to the local arm structure. 
     The luminosity functions (LFs) have been constructed for both
     clusters down to $V \sim 20$ together with their initial mass
     functions (IMFs) for stars with masses above $M \sim 1-1.5
     M_{\sun}$. The IMF slopes for the most massive bins yielded
     values of $x \approx 1.3$ for NGC~2580 and $x \approx 2$ for
     NGC~2588. In the case of this latter cluster we found evidence of
     a core-corona structure produced probably by dynamical effect. 
     In the main sequences of both clusters we detected gaps, which
     we suggest could be real features.

  }

\authorrunning{Baume et al.}
\titlerunning{NGC 2580 and NGC 2588}

\maketitle
%

\section{Introduction}

$\hspace{0.5cm}$ This study is part of a long term project based on
broad band CCD photometric observations of open clusters, aimed at
investigating the spiral structure and the star formation history in
the third Galactic quadrant; it is of particular interest to recognise 
the shape of the Vela-Puppis region (Moitinho 2001, 2002; Giorgi et
al. 2002).

Open clusters are excellent targets to be used as natural
laboratories in order to understand several issues related to the
structure, the chemical population, the dynamical evolution and
the stellar formation processes in the Galaxy. In this study, we
pay attention to the open clusters NGC~2580 and NGC~2588.  Both
objects have been poorly studied up to now (identification
and eye estimates of their angular size and richness). Preliminary
results of NGC~2588 using part of the present data were given
earlier by Moitinho (2001, 2002).

Both clusters are situated in a zone of the Galaxy where, according
to Neckel \& Klare (1980), the visual absorption rises monotonically
up to about 1~kpc from the Sun and is then nearly constant with a
value $\sim 0.5^{m}$ up to a distance of 2 kpc. At this distance, the
absorption makes a significant jump reaching $2^{m}$, a value that
apparently persists up to as far as 5 or 6 kpc. The coordinates of the
clusters taken from Dias et al. (2002) are shown in Table~1.

\begin{table}
\caption{Central coordinates of the observed objects.}
\fontsize{8} {10pt}\selectfont
\begin{center}
\begin{tabular}{ccccc}
\hline
\hline
\multicolumn{1}{c} {$Name$} &
\multicolumn{1}{c} {$\alpha_{2000}$}  &
\multicolumn{1}{c} {$\delta_{2000}$}  &
\multicolumn{1}{c} {$l$} &
\multicolumn{1}{c} {$b$} \\
\hline
NGC~2580 & 08:21:28.0 & -30:18:00.0 & $249.90^{\circ}$ & $+3.69^{\circ}$ \\
NGC~2588 & 08:23:10.0 & -32:58:30.0 & $252.29^{\circ}$ & $+2.45^{\circ}$ \\
\hline
\hline
\end{tabular}
\end{center}
\end{table}

In this paper we present the first estimate of the clusters'
fundamental parameters: distance, reddening, size and age, as well as
their Luminosity and Initial Mass Functions. All these parameters will
provide us with valuable clues for clarifying the issues enumerated above.
For this purpose, $V(I)_C$ observations in the direction of NGC~2580
and $UBV(RI)_C$ in the case of NGC~2588 were performed. These optical
measurements were complemented with infrared data taken from the 2MASS
catalogue and with astrometric information available in the Tycho-2
catalogue (H{\o}g et al. 2000) for the brightest cluster stars.

\begin{figure}
\centering
\includegraphics[width=8cm]{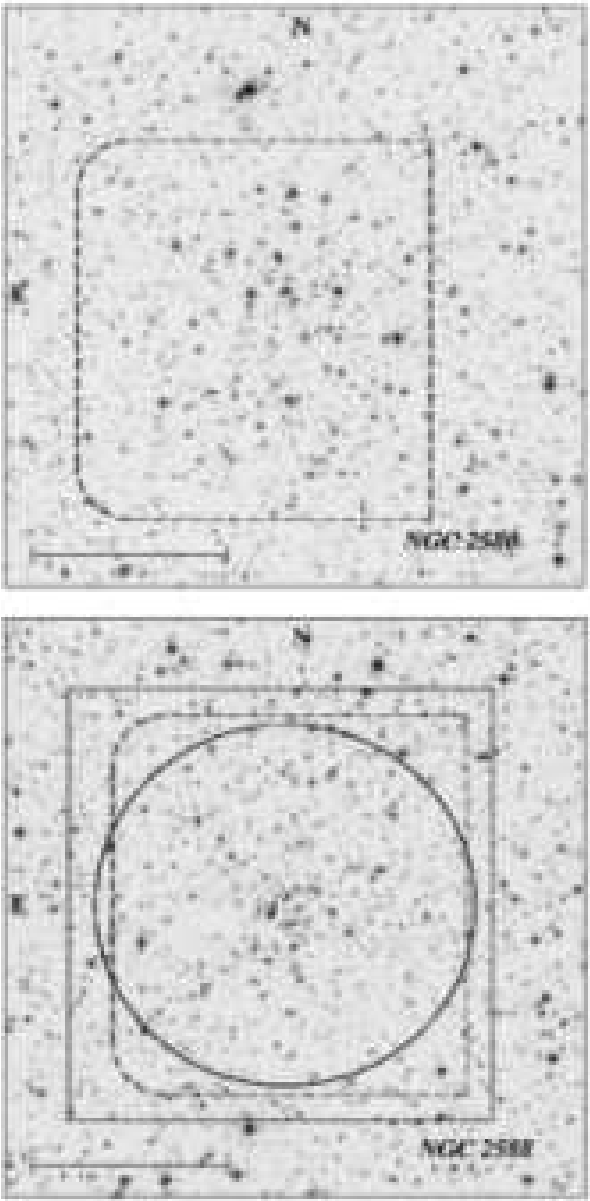}
\caption{Second generation Digitized Sky Survey (DSS-2), red filter images of
the fields centred on NGC~2580 and NGC~2588 (see Table~1). Circles, dashed
and dotted lines indicate areas covered by CASLEO, ESO and CTIO respectively.}
\end{figure}

The plan of this study is as follows: In Sect.~2 we briefly present
the observations and data reduction procedure. In Sect.~3 we describe
the methods used in the determination of the basic cluster parameters.
In Sect.~4 we present our results for NGC~2580 and NGC~2588. Sect.~5
is dedicated to a discussion and, in Sect.~6, we end with some
remarks.

\section{Data set}

\subsection{Observations}

$\hspace{0.5cm}$ Our data set comes from CCD photometric observations
carried out in the course of several runs on different telescopes:

\begin{table}
\fontsize{8} {10pt}\selectfont
\caption{Exposure times used at ESO and CASLEO together with the adopted
calibration equations and coefficients.}
\begin{center}
\begin{tabular}{lrrrrr}
\hline
\hline
\multicolumn{6}{c}{Exposure times~[sec]} \\
\hline
& \multicolumn{3}{c}{CASLEO} & \multicolumn{2}{c}{ESO} \\
\multicolumn{1}{c}{Field} & $U$ & $B$ & $V~~$ & $V$ & $I~~$ \\
\hline
 NGC~2580       &    - &    - &    -~~ & ~~300 & 240~~ \\
                &    - &    - &    -~~ &    15 &  15~~ \\
                &    - &    - &    -~~ &     1 &   1~~ \\
[0.5 ex]
 NGC~2588       & 2700 & 1800 &  600~~ &   300 & 240~~ \\
                &    - &  180 &   60~~ &    15 &  15~~ \\
                &    - &   60 &   10~~ &     1 &   1~~ \\
[0.5 ex]
\hline
PG 0918+029     &  600 &   90 &   20~~ &    15 &  10~~ \\
[0.5 ex]
PG 0942-029     &    - &    - &    -~~ &    30 &  20~~ \\
[0.5 ex]
SA 098-562      &    - &    - &    -~~ &    15 &  10~~ \\
[0.5 ex]
SA 101-424      &    - &    - &    -~~ &    30 &  20~~ \\
[0.5 ex]
Rubin 149       &  600 &   30 &   10~~ &     - &   -~~ \\
[0.5 ex]
\hline
\hline
\multicolumn{6}{c}{Calibration equations} \\
\hline
\multicolumn{1}{r}{CASLEO} & \multicolumn {5}{l}{$u = U + u_1 + u_2 (U-B) + u_3 X$}             \\
                           & \multicolumn {5}{l}{$b = B + b_1 + b_2 (B-V) + b_3 X$}             \\
                           & \multicolumn {5}{l}{$v = V + v_{1bv} + v_{2bv} (B-V) + v_{3bv} X$} \\
[1 ex]
\multicolumn{1}{r}{ESO}    & \multicolumn {5}{l}{$v = V + v_{1vi} + v_{2vi} (V-I) + v_{3vi} X$} \\
                           & \multicolumn {5}{l}{$i = I + i_1 + i_2 (V-I) + i_3 X$}             \\
[0.5 ex]
\hline
\hline
\multicolumn{6}{c}{Calibration coefficients} \\
\hline
\multicolumn {3}{c}{CASLEO (25/01/1996)}          & \multicolumn {3}{c}{ESO}                          \\
[0.5 ex]
\multicolumn {3}{l}{$~~~~u_1 = +7.595 \pm 0.025$}     & \multicolumn {3}{l}{$v_{1vi} = -0.560 \pm 0.023$} \\
\multicolumn {3}{l}{$~~~~u_2 = -0.230 \pm 0.041$}     & \multicolumn {3}{l}{$v_{2vi} = -0.058 \pm 0.023$} \\
\multicolumn {3}{l}{$~~~~u_3 = +0.49$}                & \multicolumn {3}{l}{$v_{3vi} = +0.135$}           \\
[0.5 ex]
\multicolumn {3}{l}{$~~~~b_1 = +2.611 \pm 0.006$}     & \multicolumn {3}{l}{$i_1 = -0.258 \pm 0.066$}     \\
\multicolumn {3}{l}{$~~~~b_2 = -0.135 \pm 0.010$}     & \multicolumn {3}{l}{$i_2 = -0.063 \pm 0.070$}     \\
\multicolumn {3}{l}{$~~~~b_3 = +0.27$}                & \multicolumn {3}{l}{$i_3 = +0.048$}               \\
[0.5 ex]
\multicolumn {3}{l}{$~~~~v_{1bv} = +1.406 \pm 0.004$} & \multicolumn {3}{l}{}                             \\
\multicolumn {3}{l}{$~~~~v_{2bv} = +0.054 \pm 0.006$} & \multicolumn {3}{l}{}                             \\
\multicolumn {3}{l}{$~~~~v_{3bv} = +0.12$}            & \multicolumn {3}{l}{}                             \\
[0.5 ex]
\hline
\hline
\end{tabular}
\end{center}
\end{table}

\begin{table*}
\caption{Some of the brightest stars of NGC 2580 and NGC 2588}
\fontsize{8} {10pt}\selectfont
\begin{center}
\begin{tabular}{crrrrcccccl}
\hline
\hline
\multicolumn{11}{c} {$NGC 2580$}          \\
\hline
\multicolumn{1}{c} {$\#$}            &
\multicolumn{1}{c} {$X[^{\prime\prime}]$} & \multicolumn{1}{c} {$Y[^{\prime\prime}]$} &
\multicolumn{1}{c} {$\alpha_{2000}$}      & \multicolumn{1}{c} {$\delta_{2000}$}      &
\multicolumn{1}{c} {$V$}                  &
\multicolumn{1}{c} {$V-I$}                &
\multicolumn{2}{c} {$Remarks$}            &                                           & \\
\hline
   3 &   -1.0 &  112.0 &  8:21:28.0 & -30:16:07.9 & 11.44 & 0.45 & $lm$ & \multicolumn{3}{l}{TYC 7122 385 1  } \\ 
   5 & -186.5 &   79.4 &  8:21:42.4 & -30:16:40.6 & 11.57 & 0.38 & $lm$ & \multicolumn{3}{l}{TYC 7122 87 1   } \\ 
   6 &   -4.2 &  161.2 &  8:21:28.3 & -30:15:18.7 & 11.88 & 0.43 & $lm$ & \multicolumn{3}{l}{J08212833-301518} \\ 
   7 &   -2.4 &   -3.6 &  8:21:28.1 & -30:18:03.6 & 12.05 & 0.30 & $lm$ & \multicolumn{3}{l}{TYC 7122 1129 1 } \\ 
  11 &   44.8 &  154.5 &  8:21:24.5 & -30:15:25.4 & 12.55 & 0.30 & $lm$ & \multicolumn{3}{l}{J08212455-301525} \\ 
  35 &  -51.3 & -200.8 &  8:21:31.9 & -30:21:20.7 & 13.59 & 0.25 & $lm$ & \multicolumn{3}{l}{J08213194-302120} \\ 
  44 & -132.9 &   56.1 &  8:21:38.2 & -30:17:03.9 & 13.84 & 0.31 & $lm$ & \multicolumn{3}{l}{J08213825-301703} \\ 
  46 &  -73.0 &   56.5 &  8:21:33.6 & -30:17:03.5 & 13.86 & 0.39 & $lm$ & \multicolumn{3}{l}{J08213362-301703} \\ 
  61 &  -37.3 &   27.5 &  8:21:30.8 & -30:17:32.5 & 14.27 & 0.43 & $lm$ & \multicolumn{3}{l}{J08213086-301732} \\ 
  68 &  -30.9 &  -75.1 &  8:21:30.3 & -30:19:15.0 & 14.41 & 0.36 & $lm$ & \multicolumn{3}{l}{J08213037-301914} \\ 
  69 &   61.3 &  -12.7 &  8:21:23.2 & -30:18:12.7 & 14.43 & 0.36 & $lm$ & \multicolumn{3}{l}{J08212326-301812} \\ 
  75 &  -20.6 &   10.2 &  8:21:29.5 & -30:17:49.7 & 14.65 & 0.36 & $lm$ & \multicolumn{3}{l}{J08212958-301749} \\ 
  84 &  -12.3 &   30.5 &  8:21:28.9 & -30:17:29.4 & 14.80 & 0.36 & $lm$ & \multicolumn{3}{l}{                } \\ 
  86 &   33.2 &  -50.4 &  8:21:25.4 & -30:18:50.3 & 14.81 & 0.43 & $lm$ & \multicolumn{3}{l}{J08212544-301850} \\ 
  89 &   72.1 &  -44.3 &  8:21:22.4 & -30:18:44.2 & 14.86 & 0.36 & $lm$ & \multicolumn{3}{l}{J08212242-301844} \\ 
  91 & -128.4 &  -22.0 &  8:21:37.9 & -30:18:21.9 & 14.87 & 0.43 & $lm$ & \multicolumn{3}{l}{J08213786-301821} \\ 
 101 &  -45.8 &   26.2 &  8:21:31.5 & -30:17:33.7 & 14.98 & 0.44 & $lm$ & \multicolumn{3}{l}{J08213151-301733} \\ 
\hline
\hline
\multicolumn{11}{c} {$NGC 2588$}          \\
\hline
\multicolumn{1}{c} {$\#$}                 &
\multicolumn{1}{c} {$X[^{\prime\prime}]$} & \multicolumn{1}{c} {$Y[^{\prime\prime}]$} &
\multicolumn{1}{c} {$\alpha_{2000}$}      & \multicolumn{1}{c} {$\delta_{2000}$} &
\multicolumn{1}{c} {$V$}                  &
\multicolumn{1}{c} {$U-B$}                & \multicolumn{1}{c} {$B-V$}                &
\multicolumn{1}{c} {$V-R$}                & \multicolumn{1}{c} {$V-I$}                &
\multicolumn{1}{c} {$Remarks$}            \\
\hline
  26 &    1.3 &   33.3 &  8:23:09.8 & -32:57:56.6 & 13.89 &  1.10 &  1.41 &  0.76 &  1.51 & $pm$ (red)  \\
  31 &   33.4 &   26.3 &  8:23:07.3 & -32:58:03.7 & 13.97 &  0.33 &  0.36 &  0.20 &  0.46 & $lm$        \\
  38 &    4.0 &    2.2 &  8:23:09.6 & -32:58:27.8 & 14.11 &  0.29 &  0.33 &  0.18 &  0.43 & $lm$        \\
  41 &   -5.2 &  143.7 &  8:23:10.4 & -32:56:06.3 & 14.15 &  0.20 &  0.38 &  0.21 &  0.48 & $pm$        \\
  44 &  126.8 &   61.6 &  8:22:59.9 & -32:57:28.4 & 14.22 &  0.87 &  1.26 &  0.71 &  1.40 & $pm$ (red)  \\
  47 &   21.9 &   34.1 &  8:23:08.2 & -32:57:55.9 & 14.35 &  1.16 &  1.36 &  0.72 &  1.41 & $pm$ (red)  \\
  48 &  -12.0 &   21.6 &  8:23:10.9 & -32:58:08.4 & 14.35 &  0.24 &  0.32 &  0.17 &  0.42 & $lm$        \\
  53 &  -36.7 &   -6.2 &  8:23:12.9 & -32:58:36.2 & 14.43 &  0.27 &  0.32 &  0.18 &  0.44 & $lm$        \\
  56 &   -5.2 &  -19.5 &  8:23:10.4 & -32:58:49.4 & 14.51 &  0.17 &  0.31 &  0.17 &  0.43 & $lm$        \\
  64 &  -30.4 &  -61.9 &  8:23:12.4 & -32:59:31.9 & 14.61 &  0.22 &  0.40 &  0.22 &  0.49 & $pm$ (bin?) \\
  65 &    3.3 &  -90.2 &  8:23:09.7 & -33:00:00.1 & 14.66 &  0.22 &  0.31 &  0.14 &  0.38 & $lm$        \\
  69 &   51.3 &   79.2 &  8:23:05.9 & -32:57:10.8 & 14.67 &  0.22 &  0.31 &  0.18 &  0.44 & $lm$        \\
  72 &  -42.3 &   -9.9 &  8:23:13.3 & -32:58:39.8 & 14.69 &  0.89 &  1.24 &  0.69 &  1.40 & $pm$ (red)  \\
  78 &   18.8 &  -15.6 &  8:23:08.5 & -32:58:45.6 & 14.76 &  0.28 &  0.32 &  0.18 &  0.43 & $lm$        \\
  80 &   -3.2 &  -97.7 &  8:23:10.2 & -33:00:07.7 & 14.77 &  0.33 &  0.38 &  0.22 &  0.48 & $pm$ (bin?) \\
  83 &  -57.2 &   -5.8 &  8:23:14.5 & -32:58:35.8 & 14.81 &  0.28 &  0.34 &  0.19 &  0.46 & $lm$        \\
  92 &  -17.4 &   15.2 &  8:23:11.3 & -32:58:14.8 & 14.95 &  0.25 &  0.32 &  0.17 &  0.43 & $lm$        \\
  96 &  114.8 &  -17.6 &  8:23:00.8 & -32:58:47.5 & 14.99 &  0.26 &  0.35 &  0.20 &  0.48 & $lm$        \\
\hline
\hline
\end{tabular}
\begin{tabular}{c}
\begin{minipage}{16cm}
\vspace{0.1cm}
\hspace{1cm}
{\bf Note:} Table 3 that includes all the photometric measurements is 
available in an electronic version at the CDS.
\end{minipage}
\end{tabular}
\end{center}
\end{table*}

\begin{figure}
\centering
\includegraphics[width=7cm]{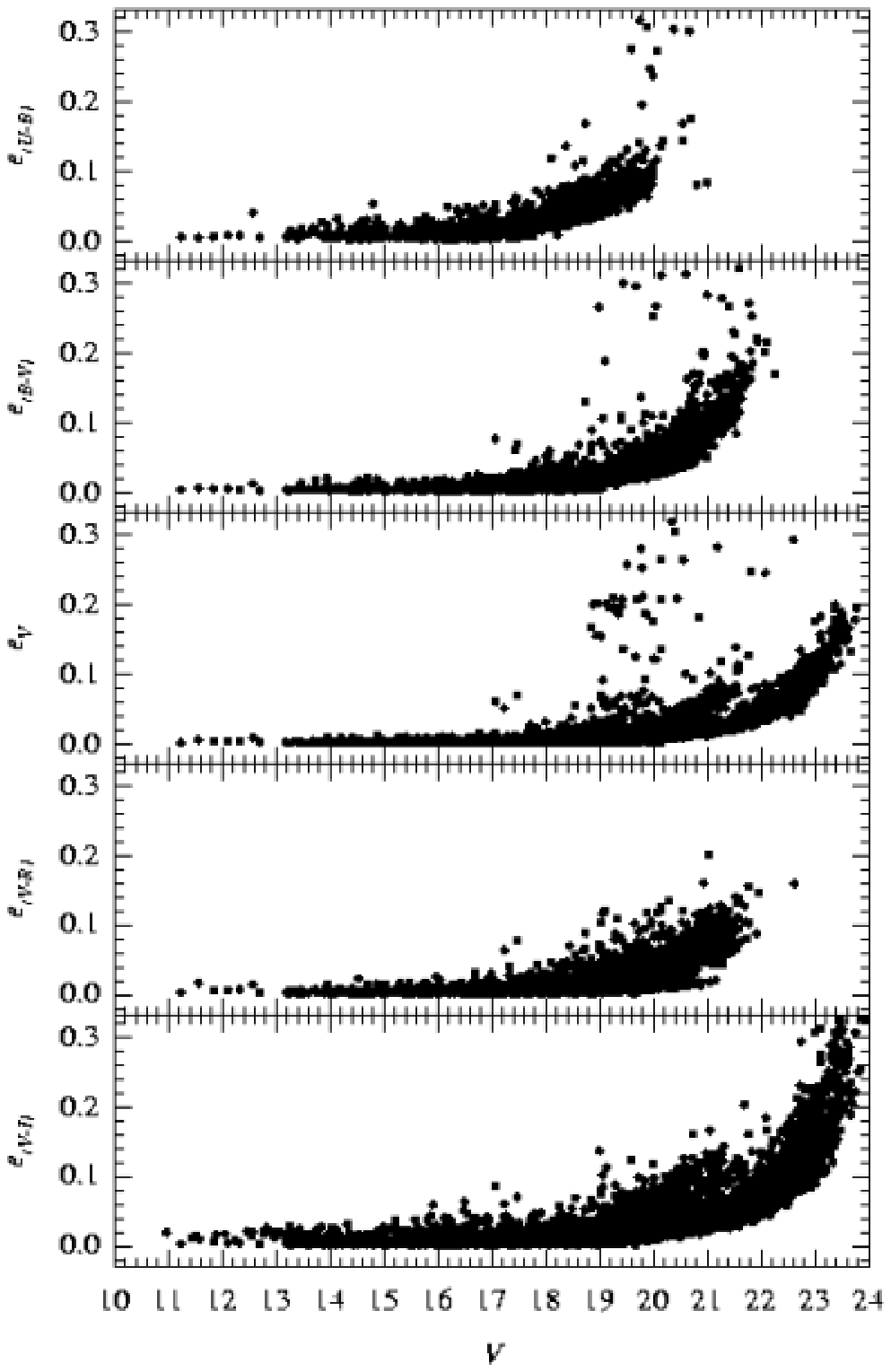}
\caption{DAOMASTER errors in the colour indexes and $V$ magnitude as a
function of $V$.}
\end{figure}

\begin{itemize}
\item[1.] The European Southern Observatory (ESO): CCD $VI$
  observations of the fields of NGC~2580 and NGC~2588 using the new
  EMMI read arm camera at the NTT in the photometric night of
  December 9, 2002 with seeing values near to $1^{\prime\prime}$. This
  new camera has a mosaic of two $2048 \times 4096$ pixels CCDs which
  samples a $9\farcm9 \times 9\farcm1$ field. The images were binned
  $2 \times 2$, which resulted in a plate scale of $0\farcs332$/pix.
\item[2.] The Complejo Astron\'{o}mico El Leoncito (CASLEO): In the
  night January 25, 1996 CCD $UBV$ observations of NGC~2588 were
  performed using the 215 cm telescope. The typical seeing was about
  $2^{\prime\prime}$. The CASLEO camera uses a Tek-1024 detector that
  together with a focal reducer covers a round area with a $9\farcm0$
  diameter and yields a plate scale of $0\farcs813$/pix.
\item[3.] The Cerro Tololo Inter-American Observatory (CTIO):
  $UBV(RI)_C$ CCD photometric data of the field of NGC~2588 were
  acquired with the 0.9~m telescope on the night of January 1, 1998.
\end{itemize}
Details of the observed fields and exposure times of the CASLEO
and ESO observations (including standard star frames) are
listed in Table~2 together with the expressions and coefficients
of the calibration equations. Details of the observations and the
reductions for the CTIO data are extensively given in Moitinho
(2001).  Fig.~1 shows the areas of NGC~2580 and NGC~2588 covered
on each run.

\begin{figure*}
\centering
\includegraphics[width=8.8cm]{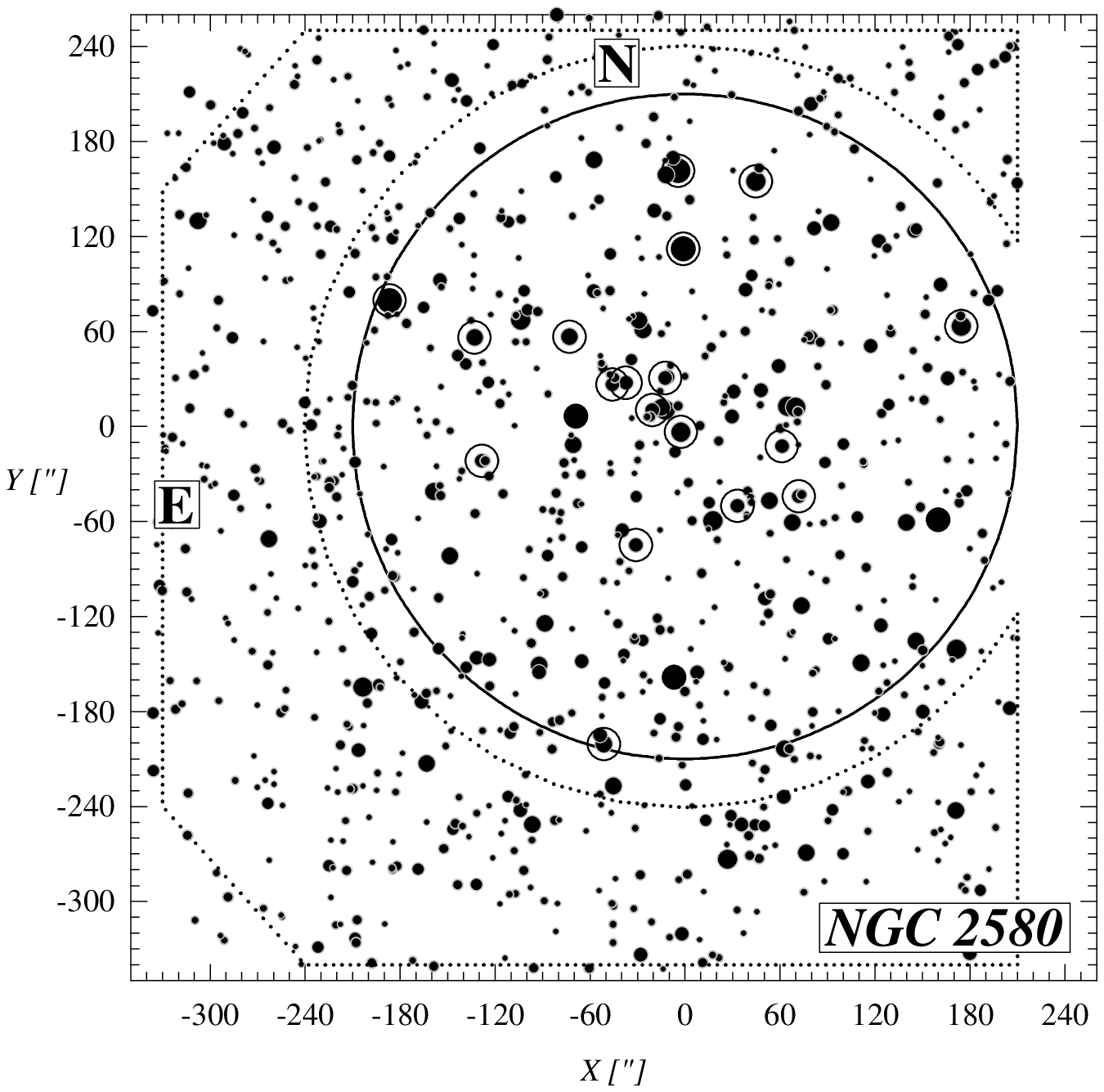}
\includegraphics[width=8.8cm]{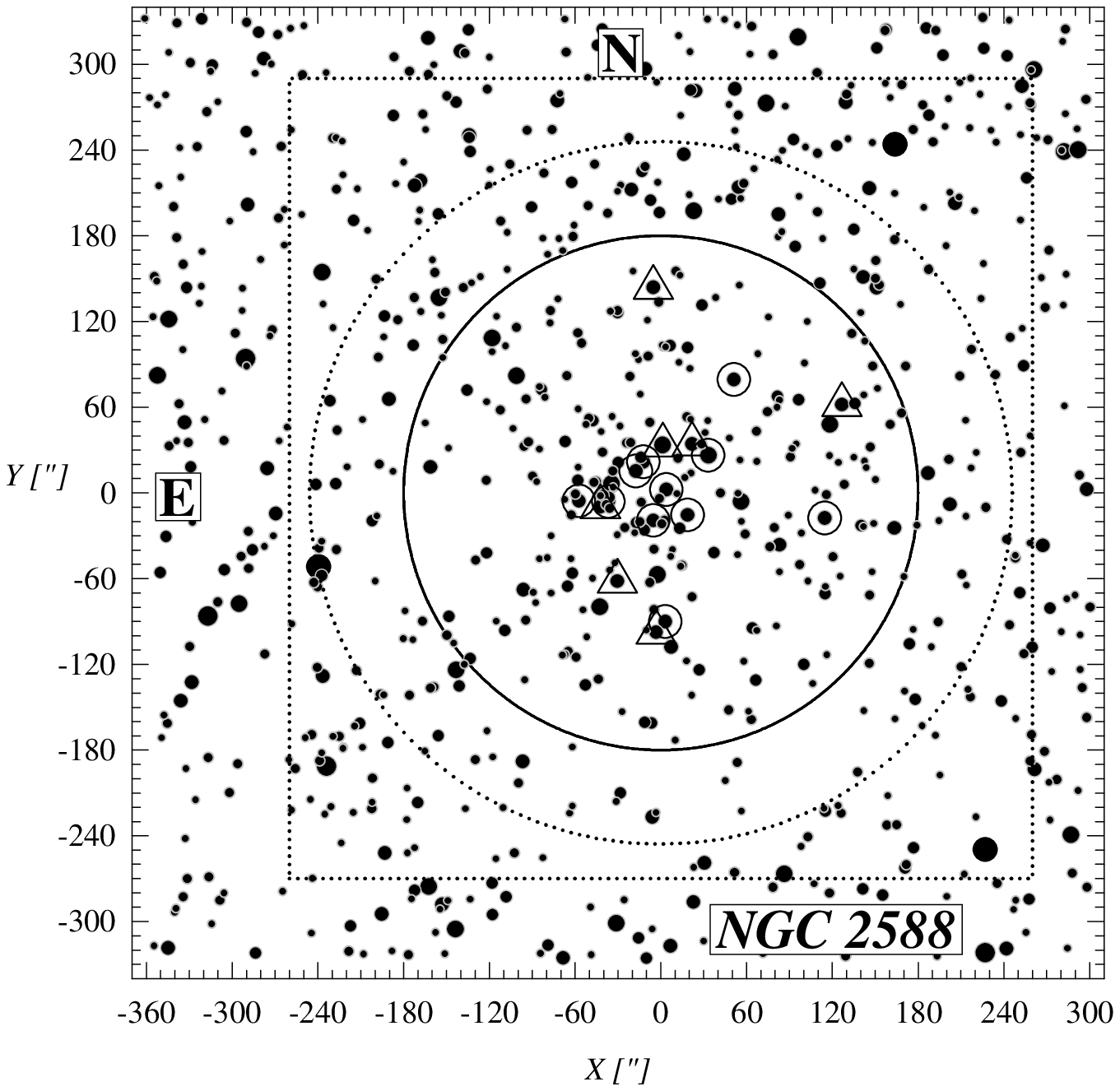}
\caption{Finding charts of the observed regions in NGC~2580 and NGC~2588
  ($V$ filter). The black solid circles, $3\farcm5$ in radius for NGC
  2580 and $3\farcm0$ for NGC 2588 indicate the adopted angular sizes
  for the clusters (see Sect.~4 and Fig. 4). The dotted lines indicate
  the areas adopted as comparison fields. Adopted likely and probable
  cluster members with $V < 15$ are indicated inside hollow circles
  and triangles, respectively. For a coordinate reference, the centre
  ($X = 0$; $Y=0$) corresponds to the cluster coordinates (see
  Table~1) and all $X-Y$ are expressed in arcseconds.}
\end{figure*}

\subsection{Reductions}

\renewcommand{\thefootnote}{\dag}

$\hspace{0.5cm}$ The ESO and CASLEO data were reduced with the
IRAF\footnote{IRAF is distributed by NOAO, which are operated by AURA
  under cooperative agreement with the NSF.} packages CCDRED, DAOPHOT,
and PHOTCAL using the point spread function (PSF) method (Stetson
1987). Calibration coefficients for ESO data were obtained through
observations of Landolt (1992) fields PG~0918+029, PG~0942-029, SA
098-562 and SA 101-424. For CASLEO observations, we used the
PG~0918+029 and Rubin 149 fields also from Landolt (1992). In all
cases, care was taken to have a good colour coverage.  Exposure
times, calibration equations and coefficients are given in Table~2,
where $UBVI$ is used to indicate standard star magnitudes, $ubvi$ for
the instrumental ones and $X$ denotes the airmass. As for the
extinction coefficients, typical values for ESO and CASLEO were
adopted. All data sets obtained from different exposure times and
different observational runs were put together using the DAOMASTER
code (Stetson 1992). The photometric error trends against the $V$
magnitude of the combined data are shown in Fig.~2. The photometric
data for some of the brightest stars of NGC~2580 and NGC~2588 are
shown in Table~3. The full table is only available electronically.

\section{Data Analysis}

\subsection{Individual stellar coordinates}

Several of the brightest stars in the observed fields have precise
astrometric measurements from the Tycho-2 catalogue (H{\o}g et al.
2000). Many other stars were also found in the 2MASS catalogue. This
information was used to compute equatorial coordinates for all the
stars in each cluster field by means of a linear transformation. The
coordinates are included in Table~3. The residuals of the
transformation were of the order of $\sim 0\farcs5$.

\begin{table}
\caption{Completeness analysis results for ESO data}
\fontsize{8} {10pt}\selectfont
\begin{center}
\begin{tabular}{crrr}
\hline
\hline
\multicolumn{1}{c} {$\Delta V$} &
\multicolumn{1}{c} {$NGC~2580$} & \multicolumn{2}{c} {$NGC~2588$} \\
&& \multicolumn{1}{c} {$Cluster$} & \multicolumn{1}{c} {$Field$} \\
\hline
15-16 & 100.0\% & 100.0\% & 100.0\% \\
16-17 &  99.8\% &  99.8\% &  99.8\% \\
17-18 &  97.1\% &  99.6\% &  99.6\% \\
18-19 &  96.7\% &  98.4\% &  99.1\% \\
19-20 &  95.3\% &  98.0\% &  97.8\% \\
20-21 &  93.6\% &  95.9\% &  96.8\% \\
21-22 &  91.1\% &  95.9\% &  95.9\% \\
22-23 &  73.2\% &  85.9\% &  87.0\% \\
23-24 &  52.2\% &  59.9\% &  65.5\% \\
\hline
\hline
\end{tabular}
\end{center}
\end{table}

\begin{figure}
\centering
\includegraphics[width=7cm]{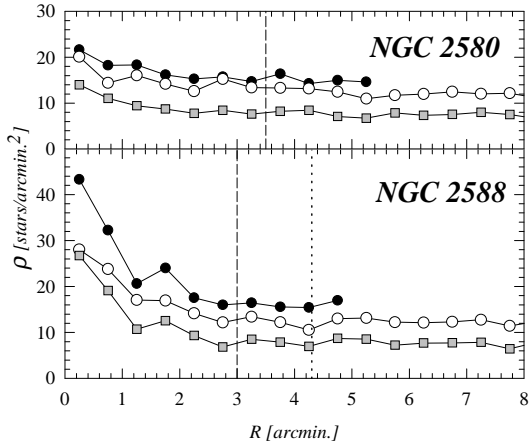}
\caption{Stellar surface density profiles in the regions of
NGC~2580 and NGC~2588 as a function of radius. The plots
represent our own data (black circles), DSS-2 images (white
circles) and data from the 2MASS catalogue (grey squares). 
Dashed lines indicate the adopted limits for the clusters. 
The dotted line is probably a better limit for NGC 2588 
(see Sect.~5.3).}
\end{figure}

\subsection{Cluster angular radius}

Deriving the radial stellar density profile for a cluster requires the
knowledge of the cluster centre. In our analysis we adopted the
cluster centres given by Dias et al. (2002, see Table~1) which were
confirmed by us, with an uncertainty of $\sim 0\farcm2$, computing the
position of the maximum of the marginal stellar distributions in both
right ascension and declination over each cluster.

The stellar density across each cluster was obtained using $a)$ our
CCD data which extend down to $V = 20$, $b)$ the corresponding DSS-2
red images of $20^{\prime} \times 20^{\prime}$ centred on each
cluster (we estimate a $V_{lim} \sim 17-18$), and $c)$ the 2MASS
infrared data $10^{\prime}$ around each cluster centre.

Stellar densities were obtained by counting the numbers of stars in a
series of concentric rings $0\farcm5$ wide divided by their respective
areas. Because of the limited area covered by our data some
rings are not complete, so we adopted the corresponding densities
as representative of the complete ring. Results are shown in Fig.~4.
Analysis of the radial density profiles and determination of the
clusters' radii will be discussed in Sect.~4.

\subsection{Cluster membership}

The most reliable assignment of cluster membership is achieved
through the analysis of proper motions and radial velocities. For the
fields of NGC~2580 and NGC~2588 most of these data are only available
for foreground stars. Although there are some kinematical data for a
few cluster stars, they have huge errors which render them useless in
a membership analysis. Therefore, to assign membership, we used the 
classical photometric criterion (e.g. Baume et al.  2003, Carraro 2002) 
by a detailed comparison of the stellar positions in all the photometric 
diagrams.

The colour-magnitude diagram (CMD) of NGC~2580 is shown in Fig.~5. 
The colour-colour diagrams (CCDs) and CMDs for NGC~2588 are shown in
Figs.~6 and 7, respectively. In Fig.~8 we present the 2MASS data CMDs
for both clusters. The CCDs of Fig.~6 and the CMDs of Figs.~5a, 7ab,
7de, 8a and 8d include all the stars inside the adopted radius for the
corresponding cluster (see Sect.~4). The diagrams shown in Figs.~5b,
7c, 7e, 8b and 8e were plotted using only stars placed around each
cluster in the regions adopted as comparison fields and denoted
$'comparison~field~1'$ (dotted areas in the finding charts of Fig.~3).
The 2MASS CMDs in Figs.~8c and 8f show alternative comparison fields
($'comparison~field~2'$, see Sect.~5.2). These latter fields were
chosen in such way that they cover equal sky surfaces as the
corresponding clusters.

The CMDs show different $V_{lim}$ (from $\sim 19.5$ in $U-B$ to $\sim
22.5$ in $V-I$) reflecting the different observational setups used in
the construction of our data set (see Sect.~2.1). The deep $V-I$ data
came from the NTT.

Both clusters exhibit a clear, blue, main sequence (MS) above 
$V \approx 16-17$ and in the case of NGC~2588 there are some bright red 
stars that could be evolved members (see Sect.~4 for details). This way, 
the individual positions in all the photometric diagrams are examined for 
each star brighter than $V \sim 17$. For stars brighter than $V \sim 16$, 
if they have coherent locations near the ZAMS and/or adopted isochrones (see 
Sect.~3.4) they are adoped as likely members ($lm$). Dimmer stars with 
magnitudes in the range $V \sim 16-17$ in the same conditions are considered 
only as probable members ($pm$). Moreover, if some stars are brighter than 
$V \sim 16$, but off the MS, they are still considered as $pm$ since their 
colour/magnitude offsets could be due to binarity or evolutionary effects 
(see Sects.~4.1 and 4.2). Additionally, the number of stars in each 
magnitude bin is checked for agreement with the counts that are obtained from 
the sustraction between the $'cluster~fields'$ and the corresponding 
$'comparison~fields'$ (see Sect.~3.5). At fainter magnitudes, contamination 
due to the Galactic field population (compare with CMDs of the 
$'comparison~fields'$) becomes severe, preventing us from easily identifying 
cluster members.

\begin{figure*}
\centering
\includegraphics[height=9cm]{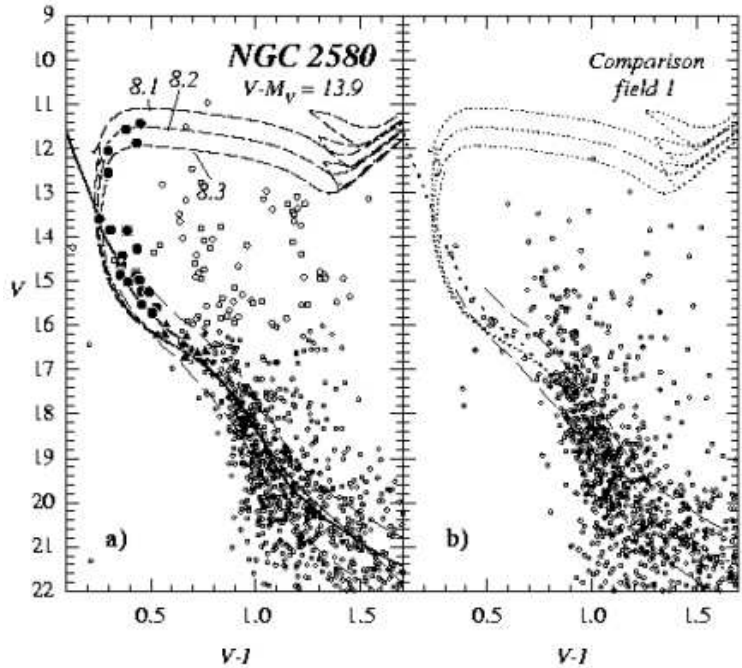}
\caption{Colour-magnitude diagrams (CMDs) of stars located inside the adopted
  radius of NGC~2580 and in the adopted comparison field.  The symbols
  have the following meaning: black circles are adopted likely member
  stars ($lm$), black triangles are probable member stars ($pm$),
  white circles are non-member stars ($nm$), and small hollow circles
  are stars without any membership assignment.  The solid line and
  dashed curves are the Schmidt-Kaler (1982) empirical ZAMS and
  isochrones from Girardi et al. (2002) respectively, corrected for
  the effects of reddening and distance. The adopted apparent distance
  modulus is $V-M_{V} = 13.9$ ($V-M_{V} = V_{0}-M_{V} + 3.1 E_{B-V}$,
  see Sect.~4). The numbers indicate $\log(age)$. Dotted lines showed
  on the comparison fields have the same meaning as the curves in the
  other panels. Long dashed lines in the $V$ vs. $V-I$ diagrams are
  the adopted envelopes used to compute the LF (see Sect.~3.5).}
\end{figure*}

\begin{figure*}
\centering
\includegraphics[height=9cm]{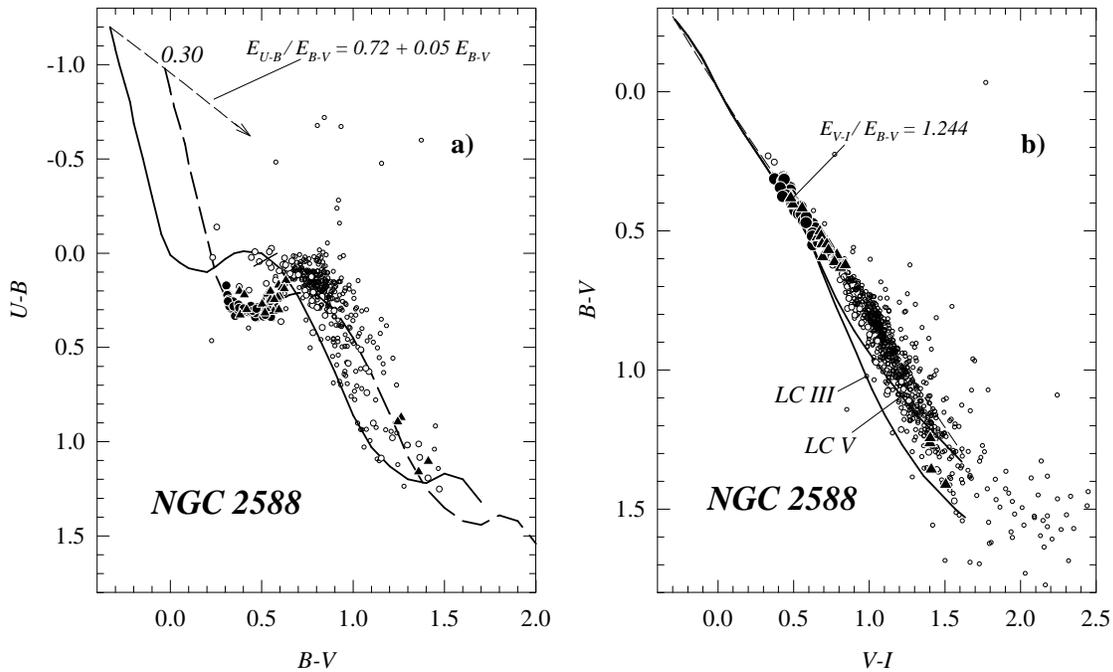}
\caption{Colour-colour diagrams (CCDs) of stars located inside the adopted
  radius of NGC~2588 {\bf a)} $U-B$ vs. $B-V$ diagram. Symbols as in
  Fig.~5. The solid line is the Schmidt-Kaler (1982) ZAMS, whereas the
  dashed one is the same ZAMS, but shifted by the adopted colour
  excess (see Sect.~4). The dashed arrow indicates the normal
  reddening path. {\bf b)} $B-V$ vs. $V-I$ diagram. Symbols and lines
  have the same meaning as in Fig.~5.}
\end{figure*}

\begin{figure*}
\centering
\includegraphics[height=9cm]{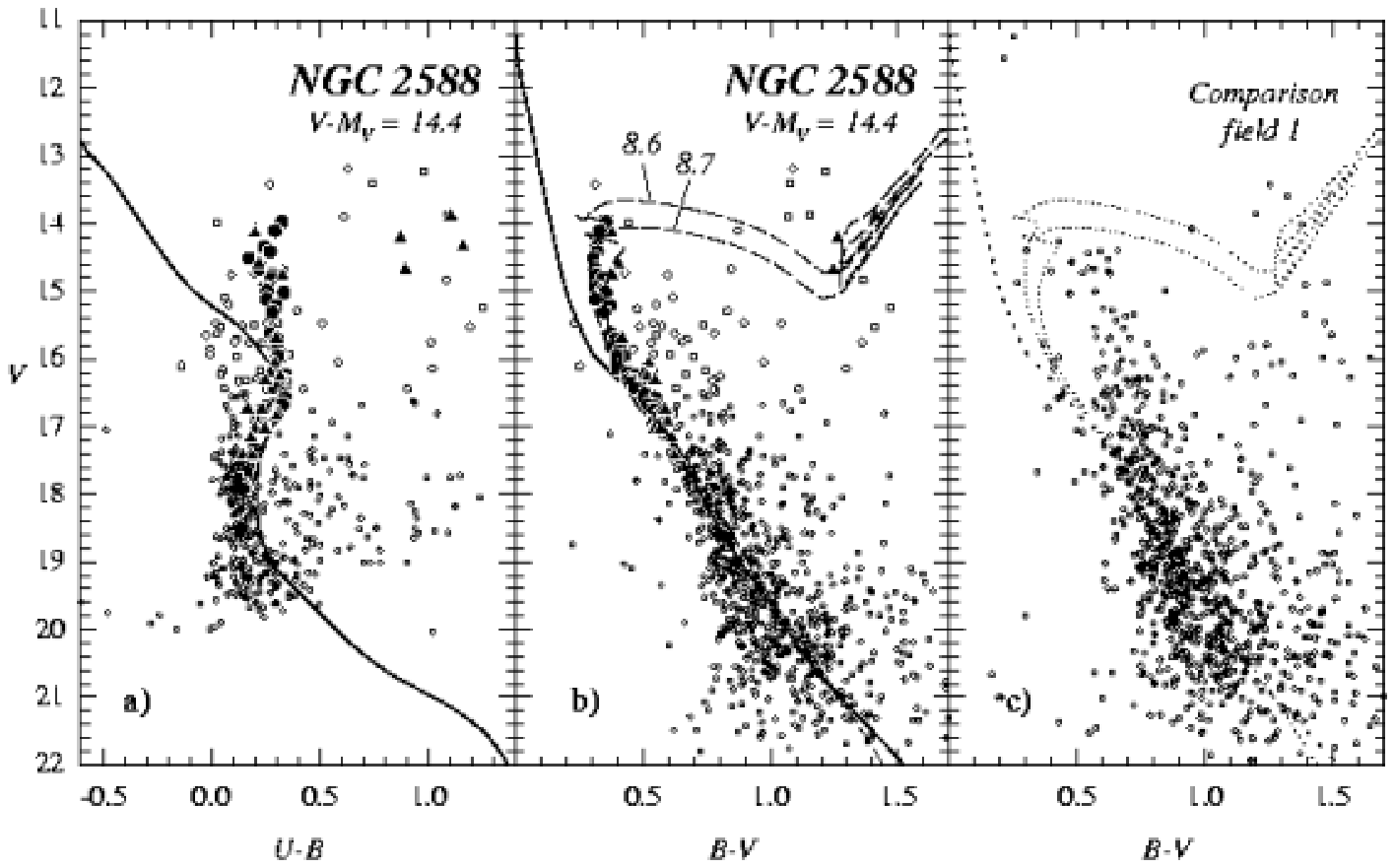}
\includegraphics[height=9cm]{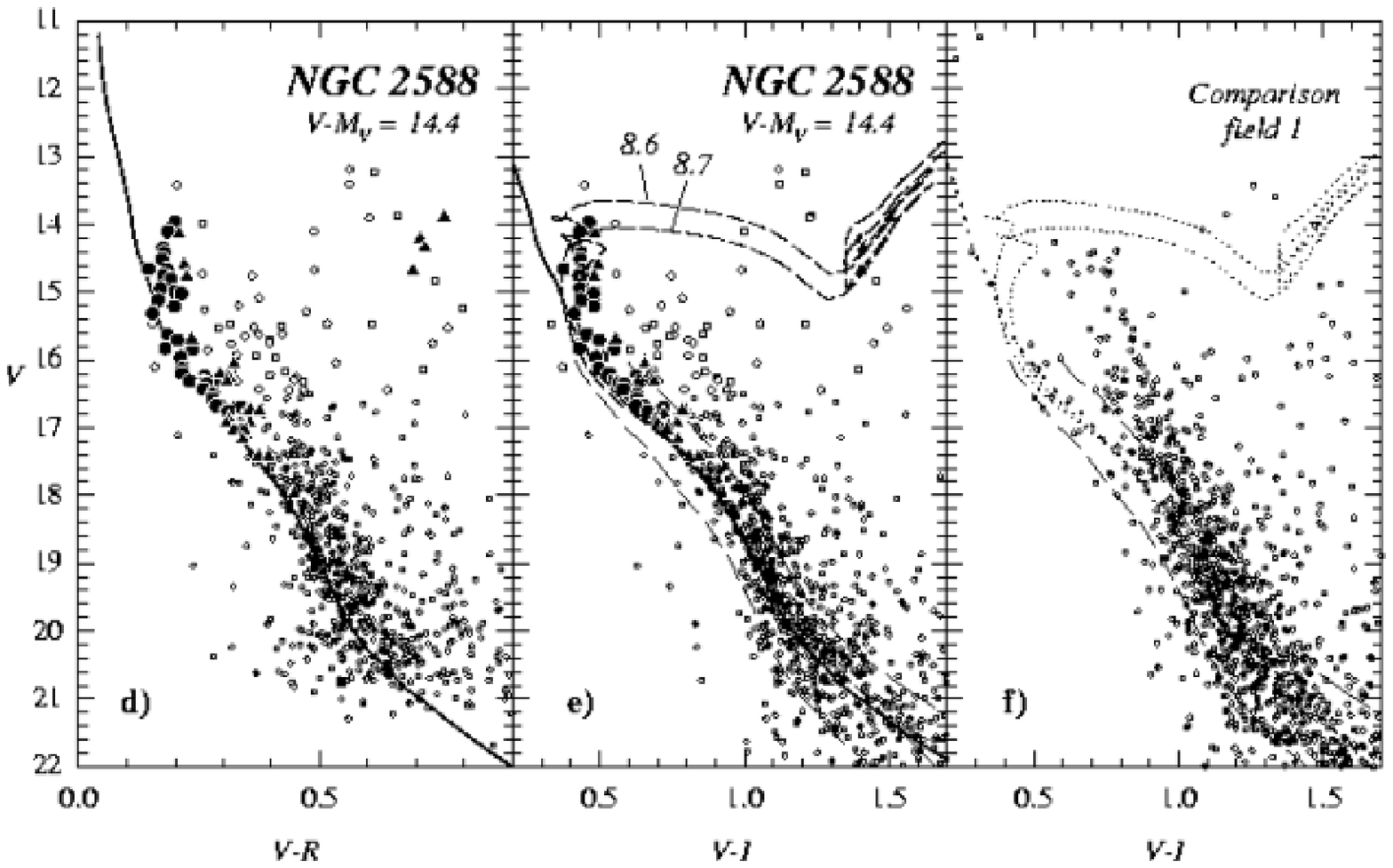}
\caption{CMDs for NGC~2588 and its comparison field. Symbols and lines have
the same meaning as in Fig.~5.}
\end{figure*}

\subsection{Cluster parameters}

Following Moitinho (2001), it appears that the third Galactic
quadrant is characterised by a reddening slope and a ratio of the
total to selective absorption ($R = A_V / E_{B-V}$) that can be
considered normal. Since the CCDs of NGC~2588 shown in Fig.~6
agree with the first assumption we adopted the $E_{U-B} / E_{B-V}
= 0.72 + 0.05~E_{B-V}$ and $E_{V-I} / E_{B-V} = 1.244$ (Dean et
al. 1978) ratios and $R = 3.1$  to shift the Schmidt-Kaler (1982)
ZAMS over the CCDs and CMDs in order to obtain the corresponding
colour excess and distance modulus for each cluster using the
traditional ZAMS fitting procedure.

To estimate the clusters' ages, isochrones derived by Girardi et
al. (2000) computed for solar metallicity, mass loss and
overshooting were superposed onto the respective de-reddened CMDs.
We looked for the ones producing the best fit over the stars
placed along the MS and also over the possible red
clump associated to each cluster. Also, from the location of
adopted cluster member stars on the CCDs along the shifted ZAMS
(dashed curves), we infer the spectral type corresponding to the
earliest MS star and we use it as another indicator of
the cluster ages.

\subsection{Cluster LF and IMF}

In this study, we present the Luminosity Function (LF) of each
cluster - the distribution of stars over the luminosity range - in
magnitude bins $1^m$ wide. The IMF will be represented as the
corresponding distribution of original ZAMS stellar masses in
logarithmic bins.

To derive each LF, we first computed the apparent magnitude 
distribution of cluster stars and then shifted this distribution by 
using the derived cluster distance modulus. Since individual cluster
memberships are only assigned for the brightest stars (see Sect.~4),
our procedure is the following:

\begin{itemize}
\item For the brightest stars ($V < 16$): we simply counted the number of
  likely member stars plus probable members on the red clump (if it
  does exist) and built the corresponding histogram of apparent
  magnitudes.
\item For fainter stars ($V > 16$): we adopted, as boundary limits,
  the envelopes around the MS on the $V$ vs. $V-I$ plane as
  shown in Fig.~5 and 7ef, and computed the apparent magnitude
  distribution of stars located in the cluster regions and also in the
  respective comparison fields. The last two distributions were
  subtracted from each other to obtain the brightness distribution of
  cluster stars.
\end{itemize}

Since the ESO observations are the deepest ones, we used them to
compute the faint end of the LF. The completeness of these data was 
assessed by running several times the ADDSTAR task, adding 15\% of the 
number of detected stars (following the same magnitude distribution), 
and by counting the number of stars recovered by the ALLSTAR task. In 
each run, a new set of artificial stars, following a similar magnitude 
distribution, was added at random positions. A comparison of the results 
from both tasks (taking into account that not all the stars are detected 
in both the $V$ and $I$ bands) yielded the completeness percentages shown 
in Table~4. These values are used to correct the apparent magnitude 
distributions of the corresponding fields and clusters. However we must 
stress that their significance is only important at very faint magnitudes 
($V > 21$). \\

As for the computation of the IMF, we considered two cases:

\begin{itemize}
\item Stars placed along the MS; here we converted absolute
  magnitude intervals of the LFs) into stellar masses using the
  mass-luminosity relation given by Scalo (1986).
\item Stars associated to the red clumps and those evolving off the
  ZAMS; here we used the set of evolutionary tracks from Girardi et
  al. (2000) and traced back the individual path for each star to
  obtain its initial mass on the ZAMS (Baume et al. 1994).
\end{itemize}

The resulting apparent LFs and the IMFs are shown in Table~5 and Fig.~9
respectively.

\section{Results}

\subsection{NGC 2580}

As NGC~2580 (= OCL709 = C0819-301) has been catalogued as an open
cluster with a diameter of about $6^{\prime}$ (Dias et al. 2002), or
$7^{\prime}$ (Lyng\aa\ 1987), our CCD observations should completely
cover this object (see Fig.~1) and part of its surroundings. van den
Berg \& Hagen (1975) classified this cluster as 'poor' (P) and more
visible on the blue plates than on the red ones, whereas later on,
Lyng\aa\ (1987) classified it as a Trumpler class II 2 m (slight
concentration, medium range in brightness and medium richness),
confirmed by Dias et al. (2002).

The radial stellar density profile shown in Fig.~4 reveals that we are
dealing with a sparse object, in agreement with its previous
classification. According to this figure the density profile reaches
the background field star density level at $\sim 3\farcm5$. We adopted
this value as the cluster angular diameter, in total agreement with
the previous qualitative suggestion given by Lyng\aa\ (1987).

The CMD (Fig.~5) exhibits a blue sequence which we identify as the
upper part of the MS of NGC~2580, extending from $V \approx
11$ to $V \approx 17$ and from $B-V \approx 0.25$ to $B-V \approx
0.6$. Among these stars, we adopt those with $V < 16$ as likely
members ($lm$) whereas the fainter ones are considered probable
cluster members ($pm$). The resulting membership assignments are
indicated with different symbols.

After some trials, the best fits of the Schmidt-Kaler (1982) ZAMS and
the Girardi et al. (2000) isochrones to the stars adopted as $lm$ and
$pm$ is obtained for a colour excess of $E_{B-V} = 0.28 \pm 0.04$
($E_{V-I} = 0.35 \pm 0.05$), an apparent distance modulus $V-M_V =
13.9 \pm 0.2$ (errors from eye inspection) and $\log(age) \approx
8.2$.  These values place NGC~2580 at $4.0 \pm 0.3$ kpc from the Sun
with an age of about 160 Myr. According to Meynet et al. 1993, the
brightest blue stars would be late B- and early A- subgiants.

The apparent LF of NGC~2580 indicated in Table~5 presents two dips:
the first at $\Delta V = 13-14$ and the second at $\Delta V =
18-19$, confirming the visual impression given by the cluster's CMD.
In particular, the former is quite evident in both CMDs (Figs.~5
and 8a). The IMF of NGC~2580 is shown in the upper part of Fig.~9, and
its slope, from a weighted least squares fit, is $x = 1.50 \pm 0.48$
(solid line) for $7.5 < M/M_{\sun} < 1.5$ or $x = 1.31 \pm 0.45$
(dashed line) for the same range but excluding the bin related to the
gap.

\begin{table}
\caption{Apparent Luminosity Functions}
\fontsize{8} {10pt}\selectfont
\begin{center}
\begin{tabular}{ccc}
\hline
\hline
$\Delta V$ & $NGC~2580$ & $NGC~2588$ \\
\hline
12-13 & ~~3 & - \\
13-14 & ~~2 & ~~2 \\
14-15 & ~~3 & ~13 \\
15-16 & ~~9 & ~10 \\
16-17 & ~10 & ~16 \\
17-18 & ~~7 & ~18 \\
18-19 & ~~3 & ~~7 \\
19-20 & ~13 & ~11 \\
20-21 & ~17 & -24 \\
21-22 & ~20 & -40 \\
\hline
\hline
\end{tabular}
\end{center}
\end{table}

\begin{figure*}
\centering
\includegraphics[height=7.5cm]{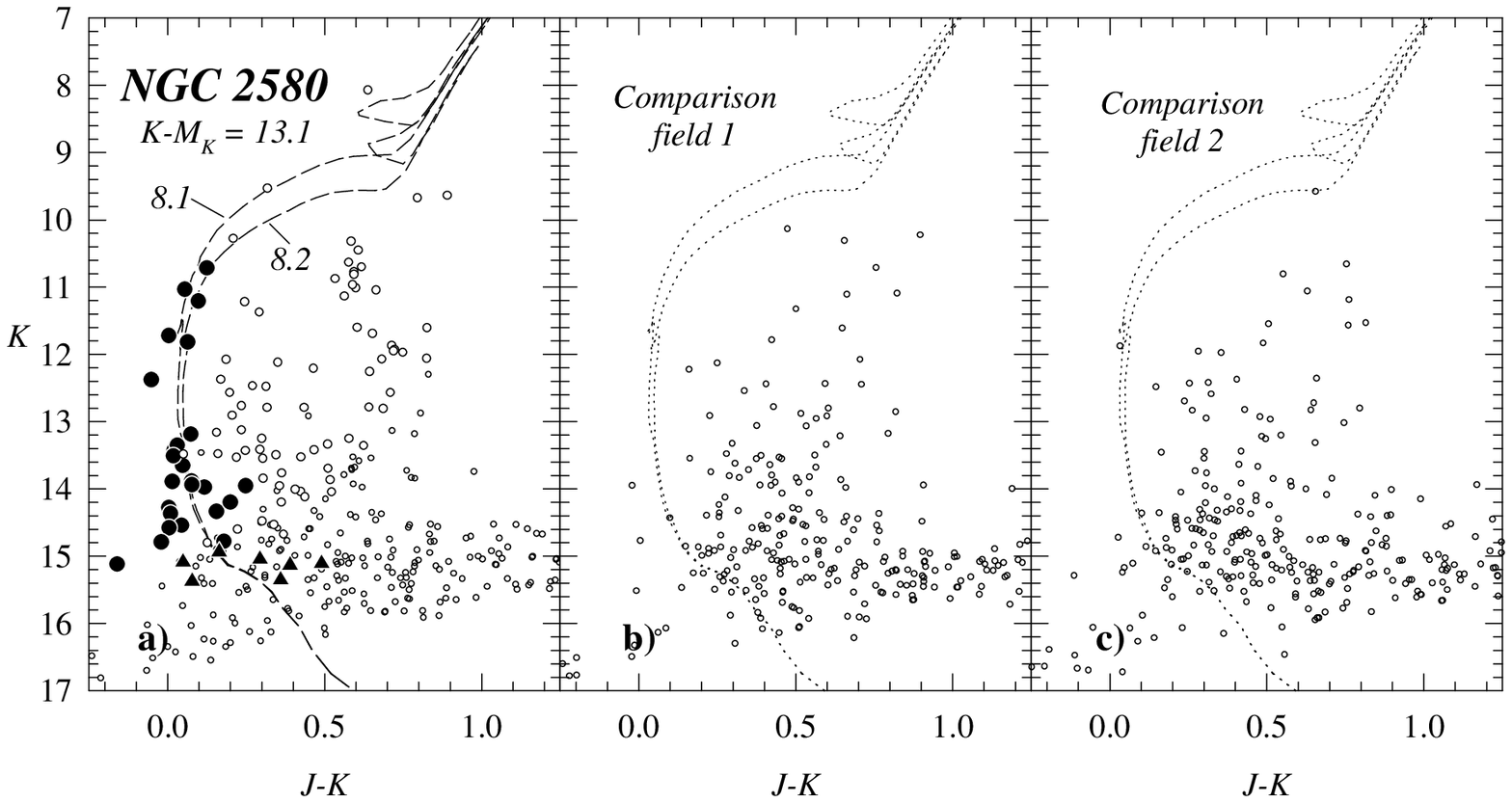}
\includegraphics[height=7.5cm]{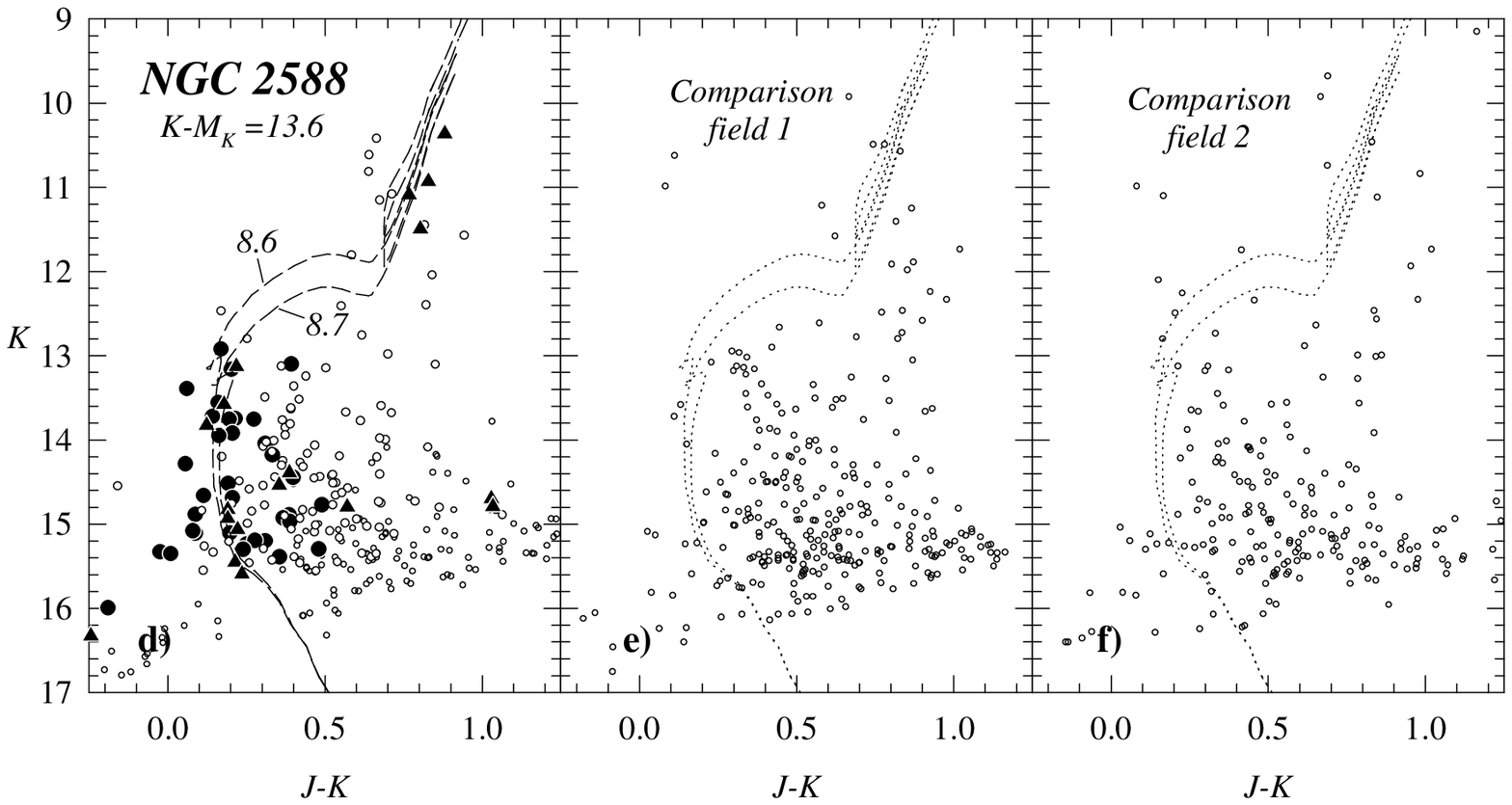}
\caption{$K$ vs. $J-K$ diagrams for stars inside the adopted radius for
  NGC~2580 and NGC~2588 and their comparison fields. Symbols are as in
  Fig.~5. Dashed curves are the isochrones from Girardi et al. (2002),
  corrected for reddening and for the adopted apparent distance moduli
  ($K-M_K = V_O-M_V + (3.1 - 2.8)~E_{B-V}$, see Sect.~4). The numbers 
  indicate $\log(age)$. The dotted lines on the comparison fields have 
  the same meaning.}
\end{figure*}

\subsection{NGC 2588}

NGC~2588 (= OCL715 = C0821-328) is a compact group of faint stars
within a diameter of about $1\farcm5$ (Dias et al. 2002) or $2^{\prime}$
(Lyng\aa\ 1987). So, at first view, our CCD observations should
completely cover this object (see Fig.~1) and also a representative
part of the field around it. According to van den Berg \& Hagen
(1975), the richness of this cluster is 'moderate' (M) and it is more
noticeable on the blue plates than on the red ones. From the Lyng\aa\
(1987) classification, this object is a Trumpler II 1 p (slight
concentration, most stars of nearly the same brightness and poor).

Fig.~4 exhibits a well shaped radial stellar density profile for this
cluster, with the highest star density inside the first $2^{\prime}$
from the centre and reaching the field star density level at
$\sim 3^{\prime}$; this value is the one we adopted as the cluster
radius. This way, the obtained diameter is larger than the one
qualitatively suggested by Lyng\aa\ (1987) and Dias et al. (2002) that
seem to only represent the core of this object.

Inspection of the different photometric diagrams (Figs.~6 and 7)
allows us to identify the stars that compose a cluster sequence
extending from $V \approx 14$ to $V \approx 17$ and from $B-V \approx
0.25$ to $B-V \approx 0.6$. As in the case of NGC~2580, we adopt those
with $V < 16.5$ as likely members ($lm$) whereas the fainter stars and
those slightly above the sequence are considered probable cluster
members ($pm$). Besides, the comparison to the best isochrones fits 
shown in Figs.~7b and 7e allows the identification of some bright red 
stars (without a counterpart in the corresponding CMDs of the comparison
field, Figs.~7c and 7f) that constitute the red clump associated to
this cluster. However, since this region of the CMDs is affected by a
certain degree of contamination by field stars, we add these candidate
red-clump stars to our list of probable cluster members ($pm$).

\begin{figure}
\centering
\includegraphics[width=7cm]{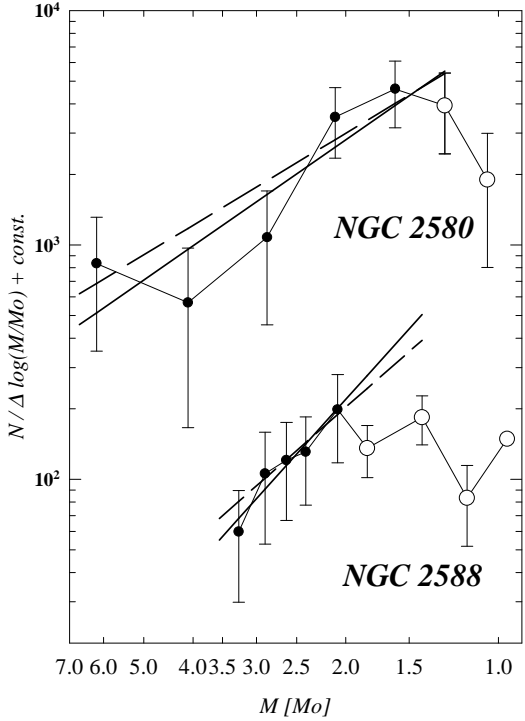}
\caption{Initial Mass Function (IMF) of NGC~2580 and NGC~2588. Error bars
  are from Poisson statistics. The weighted least square fittings for
  the more massive bins are indicated by solid and dashed right
  lines (open symbols indicate bins not used in the fits. See text
  for details). For display purposes the IMFs are shifted by an 
  arbitrary constant.}
\end{figure}

The colour excess and apparent distance modulus for this cluster,
according to the best fit to the stars adopted as $lm$ and $pm$ (see
Figs.~7a-b and 7d-e), are $E_{B-V} = 0.3 \pm 0.02$ and $V-M_{V} = 14.4
\pm 0.1$ (errors from eye-inspection). These results place NGC~2588 at
$4.95 \pm 0.2$~kpc.

Regarding the age of this cluster, comparison to the isochrones in
Figs.~7b and 7e reveals that NGC~2588 is about $450 \pm 50$ Myr. 
Moreover, according to the CCDs, the earliest MS stars in NGC~2588
have spectral type A0, which agrees with the age of $450$ Myr given
above (Meynet et al. 1993).

Similarly to NGC~2580, the apparent LF of NGC~2588, as indicated in
Table~5, shows a structure with two dips: the first appears at
$\Delta V = 15-16$ and the second at $\Delta V = 18-19$. The
former reveals a gap in the MS that is apparent in all the
CMDs of this cluster, but the latter can be due to over-subtraction of
the comparison field which could also explain the negative star counts
obtained at faint magnitudes (see the following section for details).
Regarding the IMF (see Fig.~9), the slopes determined from a weighted
least squares fit are: $x = 2.41 \pm 0.37$ (solid line) for $2.0 <
M/M_{\sun} < 3.5$, or $x = 1.90 \pm 0.33$ (dashed line) for $2.0 <
M/M_{\sun} < 3.0$.

\section{Discussion}

\subsection{General concepts}

NGC~2580 and NGC~2588 are two young/middle age open clusters located
in the third Galactic quadrant. As previously mentioned, the
determination of their fundamental parameters gives us valuable clues
for better understanding the structure of this region of our Galaxy
and its star formation history. Furthermore, it is possible to obtain
useful information concerning to the dynamical evolution of the open
clusters themselves whether internal or in relation to the Galactic disk
environment.

\subsection{The Vela-Puppis region}

Despite not being very young, NGC~2580 and NGC~2588 can be still
be used as spiral arm tracers. Putting together the data for these
two clusters with those taken from the literature (e.g. Moffat \&
Fitzgerald 1974; Moffat \& Vogt 1975; Moitinho 2002) for other
clusters placed in the same region provides a clearer picture of
the structure of the Vela-Puppis region: NGC~2588 is located, 
together with Haffner 18 (Munari et al. 1998), Haffner 19 (Munari
\& Carraro 1996), Ruprecht 32 (Moffat \& Vogt 1975), Ruprecht 44
(Moffat \& Fitzgerald 1974), NGC 2453 (Moitinho 2002) and Ruprecht
55 (Bosch et al. 2003), in a place of the Galaxy that can be 
considered an extension of the Perseus arm. As for NGC~2580, it is 
located near Haffner 16 (Moitinho 2002). These two clusters are 
located closer to the Sun than those mentioned above, in the 
inter-arm region beyond the third quadrant extension of the local 
arm. 

\subsection{Cluster IMFs}

It is evident from Fig.~9 that the IMFs of NGC~2580 and NGC~2588 
cannot be represented by a unique power law along their entire mass
range (as in NGC~4815, Prisinzano et al. (2001)). The slopes for both
clusters, derived from the most massive bins, are given at the end of
Sects. 4.1 and 4.2 respectively. 

Several studies of different clusters suggest that the spread in IMF
slopes for stellar masses above $\approx 1 M_{\sun}$ is quite large
but we do not have a clear idea yet of its true meaning: a) the IMF 
is not universal or b) there are intrinsic mistakes in its computation
(Scalo 1998). Despite that, the slope values assumed as normal are $x
= 1.35$ (Salpeter 1955) for field stars or $x = 1.7$ (Scalo 1998) for
$1 < M/M_{\sun} < 10$. In the case of the cluster NGC~2580 the slope
of the IMF agrees with such values but for NGC 2588 it turns out to
be quite large, though still comparable to the typical slope of open
clusters in the two oldest 'age groups' ($x \approx 2.2 - 2.3$) found
by Tarrab (1982). Anyway, as Prisinzano et al. (2001) argue, in this
kind of old objects, dynamical evolution plays a decisive role
since there has been enough time to disturb the original cluster mass
distribution. Furthermore, as discussed by several authors (e.g.
Sagar \& Richtler 1991, Kroupa et al. 1992, Scalo 1998) the effect of
unresolved binaries and mass segregation should flatten the derived
IMF.

For $M < 2~M_{\sun}$ (NGC~2588) and $M < 1.5~M_{\sun}$ (NGC~2580) the
slopes of the IMF appear inverted.  Since this could be an artifact 
caused by an improper choice of the comparison fields, which might not 
be representative of the true stellar field across the cluster surfaces,
we selected an additional comparison field for each cluster using data
from the 2MASS. These new comparison fields are placed farther
from the centre of each cluster ($\sim 7\farcm5$ for NGC~2580 and
$\sim 5\farcm5$ for NGC~2588) and are indicated as
$'comparison~field~2'$ in Fig. 8.

The CMDs of $'comparison~field~1'$ and~$2$ for NGC~2580 (Figs. 8b and
8c) are very similar, but for NGC~2588 the CMD of the
$'comparison~field~1'$ (Fig.~8e) appears more populated than that of
$'comparison~field~2'$ (Fig.~8f), especially notably in the region
of the lower MS. This evidence supports the choice of the
comparison field used for NGC~2580 as representative of the true
cluster field contamination, and its IMF reveals that it is deficient
in low mass stars. However, for NGC~2588, the rather evident
differences in the composition of the two comparison fields (Figs.~8e and
8f) leads us to consider the negative values in the faintest bins of
its LF (see Table~5) as produced by field over-subtraction. In turn, 
this over-subtraction indicates that the cluster has undergone significant 
mass segregation and that several of its members can be found beyond our 
first estimate of the cluster size ($R = 3\farcm0$); in fact it could 
reach out to $R = 4\farcm2$ (see Fig.~4). 

The observed mass segregation may be the result of dynamical
evolution or an imprint of the star formation process or the
combined effect of both. In order to check if dynamical evolution
plays an important role in the case of NGC~2588, we compute the
dynamical relaxation time ($T_E$) given by the following
expression (Spitzer \& Hart 1971):

\begin{center}
  $T_E = \frac{\large{8.9~10^5~N^{1/2}~R^{3/2}_h}} {\large{{\langle}
      M_* {\rangle}^{1/2}~log(0.4~N)}}$
\end{center}

\noindent
where $R_h$ is the radius containing half of the cluster
mass, $N$ is the number of cluster members and
${\langle}M_*{\rangle}$ is the average mass of the cluster stars.
Since $N$ and ${\langle}M_*{\rangle}$ are lower and upper limit
approximations, respectively, the obtained $T_E$ is indeed a rough
lower estimate. For NGC~2588 we adopted $R_h = 2.9$ pc
($4^{\prime}/2$ at 4.95 kpc), $N = 77$ and ${\langle}M_*{\rangle} =
1.8~M_{\sun}$ obtaining $T_E = 19.2$ Myr. Comparing this value
with the isochrone cluster age we find $Age/T_E \approx 25$, which
confirms that NGC~2588 is dynamically relaxed and the mass
segregation effect due to dynamical evolution must be important. A
similar analysis for NGC 2580 with $R_h = 2.2$ pc ($2^{\prime}$ at
3.8 kpc), $N = 50$, ${\langle}M_*{\rangle} = 1.9~M_{\sun}$ yields
$T_E = 11.6$ Myr and $Age/T_E \approx 10$. Therefore NGC~2580 is 
not as affected by dynamical effects as NGC 2588.

\subsection{Gaps in the main sequences}

A peculiarity that appears in the LFs, and also in the CMDs,
is the presence of gaps at $V \sim 13$ for NGC~2580 or $V \sim
15.5$ NGC~2588. They correspond to $M_V \sim -1$ ($M \sim
5.2~M_{\sun}$) and $M_V \sim +1$ ($M \sim 2.5~M_{\sun}$)
respectively. Although the gaps are narrower than those 
detected in other clusters such as NGC~3766 (Moitinho et al. 1997), 
NGC~2571 (Giorgi et al.  2002), Tr~1 or Be~11 (Yadav \& Sagar 2002), 
they seem to be significant. To check this, we compute 
the probability that a lack of stars in a mass interval is a result 
of random processes (see Scalo 1986 and Giorgi et al. 2002 for 
details) by using the following expression:

\begin{center}
   $P_{gap} = (M_{sup}/M_{inf})^{(-Nx)}$
\end{center}

\noindent
where $M_{sup}$ and $M_{inf}$ are the border masses of the gaps 
($M_{sup} > M_{inf}$), $x$ is the exponent of the IMF and $N$ the 
number of cluster members located above the gap (with $M > M_{sup}$).
For NGC~2580 the adopted values are $M_{sup} = 6.1$ (star \# 11), 
$M_{inf} = 3.9$ (star \# 35), $N = 5$ and $x = 1.31$. In the case 
of NGC~2588 they are $M_{sup} = 2.6$ (star \# 118), $M_{inf} = 2.3$ 
(star \# 149), $x = 1.9$ and $N = 15$ or $19$ (depending on whether 
the red clump stars are considered or not).

The resulting probabilities are $5~\%$ for NGC~2580 and $3~\%$ 
($N = 15$) or $1~\%$ ($N = 19$) for NGC~2588. The obtained values 
are low enough to suggest that the gaps in both cluster MSs may be 
real features.

\section{Conclusions}

We have presented the first detailed CCD study of the two poorly known
southern open clusters NGC~2580 and NGC~2588. In brief, we have found
that:
\begin{itemize}
\item NGC~2580 is a slightly spread-out and moderately bright open cluster
  with a radius of $3\farcm5$ and located approximately at $d =4$ kpc. This 
  cluster is the placed in the inter-arm region between the local arm and 
  the extension of the Perseus arm. We estimated a cluster reddening of 
  $E_{B-V} = 0.28$ and an age of 160 Myr.
\item NGC~2588 is more concentrated and fainter than NGC~2580. Our analysis 
  indicates that the core has a radius of about $1\farcm5 - 2^{\prime}$, 
  but the actual cluster could extend out to
  about $4\farcm2$, indicating the presence of a corona formed by
  fainter (less massive) members, probably due to dynamical evolution.
  NGC~2588 is affected by a reddening of $E_{B-V} = 0.3$. It is
  situated at about 5 kpc from the Sun, in the extension of the Perseus arm 
  and is about 450 Myr old. The CMDs and LF of this cluster exhibit a gap 
  at $V \sim 15.5$. Some red bright stars in the field are likely red-clump 
  members which deserve future spectroscopic investigation.
\end{itemize}

{\it This article is partially based on the Second Generation
  Digitized Sky Survey that was produced at the Space Telescope
  Science Institute under US government grant NAG W-2166. This study
  has also made use of: a) the SIMBAD database, operated at CDS,
  Strasbourg, France, and b) the Two Micron All Sky Survey, which is a
  joint project of the University of Massachusetts and the Infrared
  Processing and Analysis Center funded by NASA and NSF.}

\begin{acknowledgements}
  The authors acknowledge John Willis for his kind introduction at ESO
  NTT and for useful suggestions during the associated reduction of
  these data. We also thank the CASLEO staff for the technical
  support. AM acknowledges FCT (Portugal; grant PRAXIS XXI -
  BPD/20193/99).  The work of GB is supported by Padova University
  through a postdoctoral grant.
\end{acknowledgements}

\end{document}